\documentclass[lettersize,journal]{IEEEtran}
\usepackage{amsmath,amsfonts}
\usepackage{algorithmic}
\usepackage{algorithm}
\usepackage{booktabs}
\usepackage{array}
\usepackage{pdfpages}
\usepackage{url}
\usepackage{subcaption}
\usepackage[inkscapearea=page]{svg}
\usepackage{siunitx}
\usepackage{listings}
\usepackage{comment}
\usepackage{xcolor}
\usepackage{enumitem}
\usepackage{multirow}
\usepackage[font=small]{caption}
\def\BibTeX{{\rm B\kern-.05em{\sc i\kern-.025em b}\kern-.08em
    T\kern-.1667em\lower.7ex\hbox{E}\kern-.125emX}}

\title{Admittance Sensitivity-Informed Modular GP for Scalable Topology-Adaptive Power-Flow Learning}

\author{Henrique O. Caetano, Carlos Dias Maciel and Rahul K. Gupta

\thanks{Henrique O. Caetano and Carlos Dias Maciel are with the Department of Electrical and Computing Engineering, São Carlos School of Engineering, University of São Paulo, São Carlos,
SP, Brazil.}

\thanks{Rahul K. Gupta is with the Institute of Electrical and Micro Engineering, École Polytechnique Fédérale de Lausanne (EPFL), Switzerland.}

\thanks{This work was partially fomented by São Paulo Research Foundation (FAPESP), grants 2021/12220-1, 2023/07634-7 and 2024/08485-8.}
}
   
\usepackage{etoolbox}
\makeatletter
\patchcmd{\@maketitle}
  {\addvspace{0.5\baselineskip}\egroup}
  {\addvspace{-2.0\baselineskip}\egroup}
  {}
  {}
\makeatother

\begin{document}

\setlength{\textfloatsep}{1pt}
\setlength{\floatsep}{1pt}
\setlength{\intextsep}{1pt}

\maketitle

\begin{abstract}
Data-driven approaches for learning power flow models suffer from weak generalization across varying network topologies and limited computational scalability. Existing methods typically rely on training over a large set of grid topologies, which becomes impractical for large networks. 
This paper proposes a scalable and computationally efficient framework for topology-adaptive learning of power flow solutions. We propose a modular architecture consisting of bus-level Gaussian Process (GP) models, where each GP collects local features based on bus-level \textit{egonet} definition. 
The localized bus-level feature includes first-order power and admittance sensitivities, nodal injections and node degree. In addition to the modular architecture, we propose using Random Fourier Features (RFF) for feature reduction, which further enhances the computational scalability. We evaluate the effectiveness of the proposed method by simulations across multiple benchmark networks under N-1, N-2, and N-3 contingencies. Results for the PEGASE 1354 bus system under N-3 contingencies demonstrate high predictive quality, with an $R^2$ score of 0.983 and a voltage-magnitude RMSE of 0.0023 p.u. The framework maintains recall rates exceeding 98\% for detecting voltage limit violations across all test cases. Furthermore, the approach exhibits scalability, completing training and testing for the PEGASE 1354 system in 116.47 seconds while outperforming existing benchmarks in zero-shot generalization without requiring additional training samples.
\end{abstract}


\section{Introduction}

\subsection{Motivation}
Machine learning (ML) approaches, such as deep learning \cite{tiwari2024power,zhou2022deepopf,xiang2020probabilistic} and graph neural networks \cite{nakiganda2023graph,owerko2020optimal}, are increasingly utilized to AC power flow and optimal power flow (OPF) calculations, as they learn the non-linear mappings between nodal power injections and system states, and reduce reliance on computationally intensive conventional solvers. These schemes accelerate decision-making, which is necessary for real-time grid operation, particularly in modern power systems characterized by high penetrations of intermittent renewable energy sources \cite{gao2024bayesian}. However, to achieve operational reliability, these data-driven models must handle dynamic grid conditions, specifically topology changes and N-$k$ line contingencies \cite{xu2025optimal,yang2024topology,liu2022topology}. 

Most existing ML frameworks suffer from limited topology adaptiveness. A model trained on a specific grid topology typically experiences performance degradation when applied to an altered topology \cite{jia2024two}. Furthermore, current approaches face scalability bottlenecks for large-scale power grids. Training a single, system-wide ML model to encompass all possible operational scenarios and topological states in a large-scale grid requires massive datasets and incurs higher computational training times \cite{zhou2022deepopf}. To achieve topology-adaptiveness, algorithms must possess \textit{zero-shot generalizability capabilities}, defined as the ability to accurately predict system states on unseen network topologies without requiring new data samples for retraining or fine-tuning \cite{nakiganda2023graph}. 

\subsection{Literature Review}
Recent works have been dealing with the problem of topology-adaptive ML-based approaches for solving power flows. The work in \cite{chen2022meta} introduces a meta-learning approach for OPF that discovers optimal weight initialization points offline, allowing the model to adapt to new topologies with a few gradient steps. Similarly, stacked denoising auto-encoders are utilized in \cite{xiang2020probabilistic} to combine continuous injections with discrete line breaker statuses for probabilistic power flow. In \cite{jia2022convopf}, it proposes a convolutional neural networks (CNNs) which considers both the discrete network topology labels and load data. While these methods achieve generalized predictions, they require additional data samples from the new topology to fine-tune the network weights or update clustered labels, lacking true zero-shot adaptiveness.

Other approaches include graph-based architectures, such as graph neural networks (GNN) and physics-informed models to embed the grid's structural properties into the learning process. Physics-guided graph CNN \cite{gao2023physics} and topology-informed GNNs \cite{liu2022topology} integrate AC power flow equations and topology-dependent variables to improve adaptability. Other approaches embed branch features into GNNs \cite{yang2024topology}, utilize continuous admittance spaces within DNNs \cite{zhou2022deepopf}, or employ stacked attention mechanisms with 2D CNNs processing admittance matrices \cite{xu2025optimal}. Although these physics-informed and graph-based models generalize better to varying topologies, they also require fine-tuning for out-of-distribution topologies, achieving scalability and topology-adaptiveness only by retraining the model on new topologies.

Probabilistic frameworks, particularly Gaussian Process (GP) regression, have been extensively used within power grid literature \cite{tan2026gaussian}. GP models have been developed to map node voltages to power injections \cite{pareek2021framework} and construct topology transfer frameworks using historical OPF data \cite{jia2024two}. Bayesian deep neural networks have also been applied to spatio-temporal probabilistic OPF to handle renewable uncertainties \cite{gao2024bayesian}. Furthermore, multi-task vertex degree kernels within GPR frameworks have been introduced to construct probabilistic voltage envelopes using nodal sub-kernels focused on neighborhood injections \cite{pareek2025data}. Despite providing predictive confidence intervals, these probabilistic models either assume fixed network configurations, require new samples to train the target topology, or are validated on small distribution systems due to the lack of scalability.

To address scalability in larger networks containing thousands of nodes, some works have proposed decentralized and partitioned learning architectures. For example, a multi-branch DNNs 
network partitioning into subnetworks is proposed in \cite{huang2025decentralized}. A dynamic programming combined with Galerkin projection scheme is used in \cite{chen2022improved} that decomposes the global OPF into independent subproblems across multi-level networks.
While these decentralized and partitioned models achieve high computational efficiency on large-scale systems, they assume fixed network configurations and do not account for dynamic topology changes or provide uncertainty quantification.

\subsection{Key Gaps}
Table \ref{tab:tool_comparison} summarizes the capabilities of existing works across seven key attributes, highlighting three primary research gaps.

\textit{First}, while several works successfully implement topological generalizability \cite{zhou2022deepopf, xu2025optimal, liu2022topology, yang2024topology, jia2024two, chen2022meta, gao2023physics, pareek2025data}, they consistently require additional data samples from the new topology to update model parameters. No evaluated framework achieves zero-shot topology adaptiveness, which requires predicting system states on completely unobserved network configurations without retraining or fine-tuning.

\textit{Second}, the literature indicates a trade-off between topology adaptiveness and computational scalability. Models in \cite{huang2025decentralized, chen2022improved, mak2023learning} achieve scalability by partitioning the grid into multiple zones, but they assume fixed network structures and do not account for topological changes. In \cite{zhou2022deepopf, gao2024bayesian, liu2022topology}, models achieve scalability while handling varying topologies, but this is accomplished by including multiple grid configurations within the training dataset rather than performing zero-shot generalizability on unseen $N-k$ contingencies. 

\textit{Finally}, while some works explore modular, bus-level ML architectures that predict power flow variables individually for each bus \cite{pareek2025data}, these approaches lack zero-shot topology adaptiveness. They still require new data samples to update the model on unseen network configurations and have not been integrated into a scalable framework capable of predicting full power flow solutions under unobserved N-$k$ contingencies.

\begin{table}[htbp]

\centering
\caption{Comparative summary of recent works on Topology-adaptive Power-flow learning models}
\label{tab:tool_comparison}
\begin{tabular*}{\columnwidth}{@{\extracolsep{\fill}}lcccccccc}
\toprule
\textbf{Ref$^*$.} & \textbf{A1} &\textbf{A2} & \textbf{A3} & \textbf{A4} & \textbf{A5} & \textbf{A6} & \textbf{A7} \\
\midrule
\cite{zhou2022deepopf}  & \checkmark & - & - & \checkmark & \checkmark & \checkmark & - \\
\cite{xiang2020probabilistic} & - & - & \checkmark & \checkmark & - & - & - \\
\cite{gao2024bayesian} & - & - & \checkmark & \checkmark & \checkmark & \checkmark & -\\
\cite{xu2025optimal}  & \checkmark & - & - & \checkmark & \checkmark & - & - \\
\cite{yang2024topology}  & \checkmark & - & - & \checkmark & \checkmark & - & - \\
\cite{liu2022topology}  & \checkmark & - & - & \checkmark & \checkmark & \checkmark & - \\
\cite{jia2024two}  & \checkmark & - & \checkmark & \checkmark & \checkmark & - & - \\
\cite{chen2022meta} & \checkmark & - & - & \checkmark & \checkmark & - & - \\
\cite{jia2022convopf} & - & - & \checkmark & \checkmark & \checkmark & - & - \\
\cite{gao2023physics}  & \checkmark & - & - & \checkmark & \checkmark & - & - \\
\cite{pareek2021framework} & - & - & \checkmark & \checkmark & - & - & - \\
\cite{pareek2025data}  & \checkmark & - & \checkmark & \checkmark & - & - & \checkmark \\
\cite{huang2025decentralized} & - & - & - & \checkmark & \checkmark & - & \checkmark \\
\cite{chen2022improved} & - & - & - & \checkmark & \checkmark & \checkmark & \checkmark \\
\cite{mak2023learning} & - & - & - & \checkmark & \checkmark & \checkmark & \checkmark \\
\textbf{Proposed} & \checkmark & \checkmark & \checkmark & \checkmark & \checkmark & \checkmark & \checkmark  \\
\bottomrule
\end{tabular*}
\footnotesize\textsuperscript{$*$} \textbf{A1}: Topological generalizability with reduced samples; \textbf{A2}: Zero-shot Topology adaptiveness; \textbf{A3}: Uncertainty quantification; \textbf{A4}: Account for voltage phasor; \textbf{A5}: Account for generator setpoints; \textbf{A6}: Scalable for bigger systems ($> 200$ nodes); \textbf{A7}: Multiple zone partition
 
\end{table}

\subsection{Contributions}
To tackle the above-mentioned literature gaps, this work proposes a scalable and topology-adaptive power flow learning framework. The methodology achieves both generalizability and scalability by employing a modular GP architecture, where an independent predictive model is established for each individual bus. To enable prediction on unobserved grid topologies, the framework incorporates localized first-order voltage sensitivities (active power, reactive power, and admittance) as direct inputs, embedding the physical information of structural changes into the learning process. Furthermore, to overcome the computational intractability of standard GPs on large datasets, the approach approximates the covariance kernels using Random Fourier Features (RFF) and aggregates predictions through an ensemble of GP experts for efficient training and accurate prediction. The main contributions of this work are:
\begin{itemize}
    \item A modular bus-level GP architecture with localized feature extraction methodology that embeds first-order active power, reactive power, and analytical admittance sensitivities. These localized sensitivities are utilized as direct inputs to the predictive model, informing the algorithm of the physical impacts of topology changes without requiring full-dimensional grid data.
    \item An ensemble GP architecture approximated via RFFs. This approach reduces the computational complexity inherent to standard GPs, enabling scalability while achieving zero-shot generalizability on entirely unobserved grid topologies.
    \item The application of the proposed framework to predict bus-level voltage magnitudes and phase angles under unobserved N-$k$ contingencies. The uncertainty quantification of the GP ensemble is used to calculate confidence bounds, which are subsequently used to predict overvoltage and undervoltage scenarios across the system.
\end{itemize}

\begin{figure*}[t]
    \centering
    \includegraphics[width=0.85\linewidth]{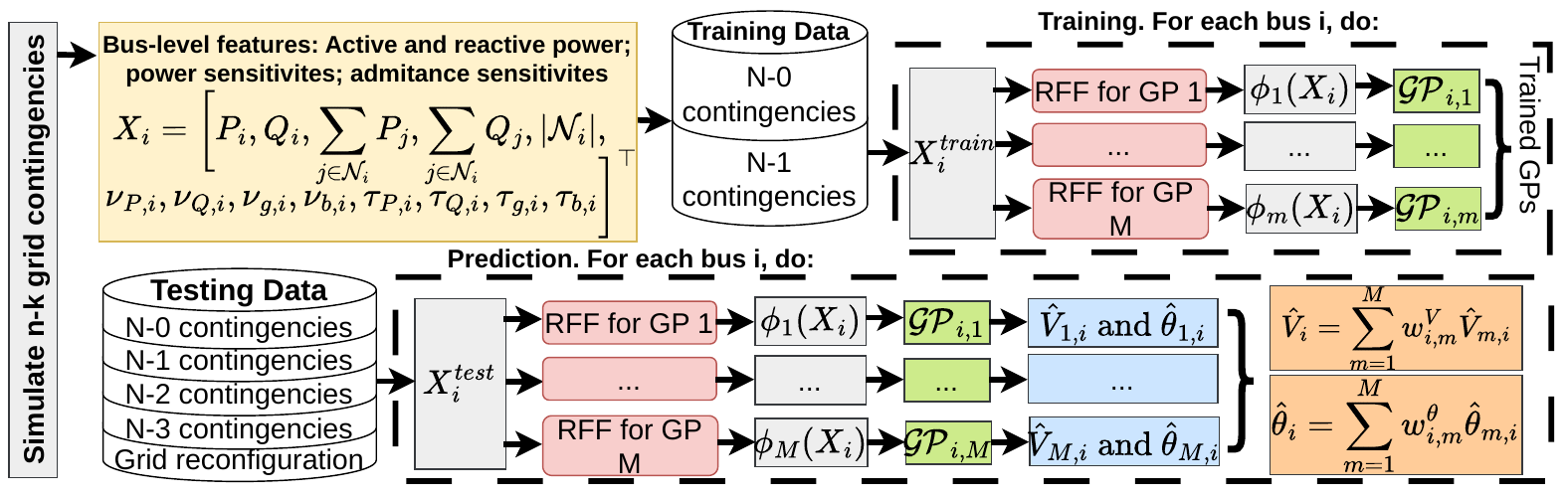}
    \caption{Flowchart of the proposed framework, which extracts bus-level operational features and physical sensitivities (yellow) that is reduced with RFF (red) to train an ensemble of $M$ Gaussian Process models per bus $i$ (green). During the testing phase, the individual predictions from each expert (blue) are combined via weight aggregation to generate the final voltage magnitude and angle predictions (orange).}
    \label{fig:flowchart}
    \vspace{-1.5em}
\end{figure*}

\section{Proposed Modular Bus-level \\Gaussian Process Model}
\label{sec:met}
We consider the nonlinear AC power flow equations, which describe the active ($P_i$) and reactive ($Q_i$) power injections at bus $i$ as functions of the bus voltages ($|V_i|\angle \theta_i$) and network admittance elements $Y_{ik}=|Y_{ik}|\angle \phi_{ik}$:
\begin{subequations}
\begin{align}
P_i &= |V_i| \sum_{k=1}^{N} |V_k| |Y_{ik}| \cos(\theta_i - \theta_k - \phi_{ik}) \\
Q_i &= |V_i| \sum_{k=1}^{N} |V_k| |Y_{ik}| \sin(\theta_i - \theta_k - \phi_{ik})
\end{align}
\end{subequations}
\noindent where $|V_i|$ and $\theta_i$ denote the voltage magnitude and phase angle at bus $i$, respectively, and $N$ is the number of buses. Solving these equations requires computing the bus voltage states, typically via the Newton-Raphson method, which becomes computationally demanding for large-scale systems.

To address this limitation, we propose a fast, scalable, and topology-adaptive learning framework for predicting bus voltage magnitudes ($|V_i|$) and angles ($\theta_i$) under unseen network topologies. Let $\mathcal{V}$ denote the set of buses with $i \in \mathcal{V}$. The framework employs a modular Gaussian Process (GP) architecture in which an independent GP model is assigned to each bus. To compensate for the absence of explicit global network coupling, each GP is augmented with localized first-order voltage sensitivities with respect to active power, reactive power, and admittance. This feature design embeds structural information into the learning process and enables adaptability to topology variations.
While prior work incorporates power flow sensitivities \cite{singh2021learning, jalali2022fast}, the proposed approach additionally integrates admittance sensitivities and introduces a modular formulation that improves scalability and generalization across varying grid topologies.

Figure \ref{fig:flowchart} illustrates the overall architecture of the proposed framework, which is structured into five sequential components. First, localized operational features and physical sensitivities are extracted directly from the AC power-flow Jacobian (Section \ref{met:egonet}). The feature input computational complexity is reduced using a Random Fourier Features (RFF) approximation (Section \ref{met:rff}). The framework then employs an ensemble learning strategy to aggregate predictions from multiple GP experts based on their marginal likelihoods (Section \ref{met:egp}). Finally, these components are unified for power flow prediction (Section \ref{met:power_flow_formulation}). In the following, we describe each of these components in detail. 

\subsection{Overview of the Bus-Level Modular GP Architecture}
For each bus $i \in \mathcal{V}$ in the network, we construct a modular architecture composed of independent, bus-level Gaussian Process (GP) models. Each bus-level model is trained to predict the local voltage magnitude $|V_i|$ and voltage angle $\theta_i$. To capture multi-scale behavior in the input–output mapping, we adopt the Ensemble GP (EGP) framework described in Section~\ref{met:egp} at each bus.
Specifically, each bus $i$ is associated with an ensemble of $M$ GP experts, where each expert is characterized by a distinct length scale. The total number of experts $M$, as well as their corresponding characteristic length scales $\{l_m\}_{m=1}^M$, are fixed \emph{a priori}. All bus-level EGP models are trained independently, enabling a fully modular and scalable learning architecture.

For a given bus $i$ and expert $m$ with length scale $l_m$, we define the target output vector as $y_i = \begin{bmatrix} |V_i|, ~\theta_i \end{bmatrix}^\top \in \mathbb{R}^2.$ Let $X_i$ denote the input feature vector associated with bus $i$, the construction of which is detailed in Section~\ref{met:egonet}. The observation model for expert $m$ is given by
\begin{equation}
    y_i = f_m\bigl(\phi_{l_m}(X_i)\bigr) + \epsilon,
\end{equation}
where $f_m$ is an unknown latent function modeled as a zero-mean Gaussian process, $f_m \sim \mathcal{GP}(0, \kappa_m),$
and $\epsilon \sim \mathcal{N}(0, \sigma_\epsilon^2 I_2)$ represents independent and identically distributed (i.i.d) Gaussian observation noise. The mapping $\phi_{l_m}(\cdot)$ denotes the RFF transformation corresponding to the length scale $l_m$ (see Section~\ref{met:rff}).
The covariance function $\kappa_m(\cdot,\cdot)$ measures the similarity between input feature vectors. Consistent with prior work \cite{polyzos2021ensemble, polyzos2021online}, we employ the Squared Exponential (SE) kernel, defined as
\begin{equation}
    \kappa_m(X_i, X'_i)
    = \sigma_f^2 \exp\Big(-\frac{\|X_i - X'_i\|^2}{2l_m^2}\Big),
\end{equation}
where $\sigma_f^2$ denotes the signal variance and $l_m$ is the characteristic length scale specific to expert $m$.

\subsection{Localized Egonet Feature Extraction and Sensitivities for Bus-level GPs}
\label{met:egonet}
The proposed bus-level modular GPs are trained using the localized operational and topological features extracted for each bus. We proposes a modified egonet \cite{polyzos2021online} feature extraction for each bus within the grid, defined as follows.

\subsubsection{Features definition} For each bus $i \in \mathcal{V}$ in the network, we define an egonet node set $\mathcal{V}_{\mathcal{E},i} = \mathcal{N}_i \cup \{i\}$, where $\mathcal{N}_i$ represent the set of buses directly connected to that bus $i$. 
We also define egonet branch sets $\mathcal{L}_{\mathcal{E},i}$ as the subset of network branches incident to at least one node within $\mathcal{V}_{\mathcal{E},i}$. These sets are used to extract features, $X_i$, for GP for bus $i$, which is composed of the local active and reactive power injections, node degree, and power and admittance sensitivities, as described below.

The \textit{first} set of input features 
include the local active and reactive power demands ($P_i$, $Q_i$) and the aggregated active and reactive powers within its egonet ($\sum_{j \in \mathcal{N}_i} P_j$, $\sum_{j \in \mathcal{N}_i} Q_j$). These features inform the GP model about behavior of the output predictors, i.e., the nodal voltage magnitudes and angles with respect to the local and neighboring loading scenarios.

The \textit{second} feature is on the node degree $|\mathcal{N}_i|$, proving the GP model information on number of neighboring connections.

The \textit{third} set of features explicitly captures the local admittance parameters of the network. This includes the raw conductance ($g$) and susceptance ($b$) values for all branches within the egonet. Similar to other topological parameters, these values are collected into vectors over the egonet branch set $\mathcal{L}_{\mathcal{E},i}$, with disconnected elements branches set to 0:
\begin{equation}
    G_i = \left[ g_k \right]_{k \in \mathcal{L}_{\mathcal{E},i}}, \quad B_i = \left[ b_k \right]_{k \in \mathcal{L}_{\mathcal{E},i}}
\end{equation}

The \textit{fourth} set of features are on the voltage sensitivities to capture the nonlinear power-flow dynamics. Specifically, the model incorporates first-order sensitivities of the voltage magnitude ($|V|$) and voltage angle ($\theta$) with respect to both nodal power injections and branch admittances. Let $\frac{\partial |V_i|}{\partial g_{j,i}}, \frac{\partial |V_i|}{\partial b_{j,i}} , \frac{\partial \theta_i}{\partial g_{j,i}} , \frac{\partial \theta_i}{\partial b_{j,i}} $ be the voltage magnitude and angle sensitivities with respect to admittance parameters, i.e., conductance ($g_{j,i}$) and suscepetance ($b_{j,i}$) between buses $i$ and $j$.
The magnitude and angle sensitivities corresponding to the admittance parameters\footnote{Detailed derivation of admittance sensitivities can be found in \cite{talkington2025differentiating, admittance_sensitivity_rahul}.} are localized by collecting their individual values over the egonet branch set $\mathcal{L}_{\mathcal{E},i}$ into vectors. To maintain a constant input dimension for the GP, if a contingency occurs in an element, its corresponding sensitivity is set to zero (an example is shown later in Sec.~\ref{sec:5bus_example}):
\begin{equation}
    \nu_{g,i} = \left[ \frac{\partial |V_i|}{\partial g_k} \right]_{k \in \mathcal{L}_{\mathcal{E},i}}, \quad \nu_{b,i} = \left[ \frac{\partial |V_i|}{\partial b_k} \right]_{k \in \mathcal{L}_{\mathcal{E},i}}
\end{equation}
\begin{equation}
    \tau_{g,i} = \left[ \frac{\partial \theta_i}{\partial g_k} \right]_{k \in \mathcal{L}_{\mathcal{E},i}}, \quad \tau_{b,i} = \left[ \frac{\partial \theta_i}{\partial b_k} \right]_{k \in \mathcal{L}_{\mathcal{E},i}}
\end{equation}

Similarly, let the voltage magnitude and angle sensitivities with respect to power injections are denoted by $\frac{\partial |V_i|}{\partial P_{j}}, \frac{\partial |V_i|}{\partial Q_{j}}, \frac{\partial \theta_i}{\partial P_{j}}, \frac{\partial \theta_i}{\partial Q_{j}}$ (extracted from the corresponding rows of the inverse Jacobian $J^{-1}$). The localized features are similarly evaluated by collecting the individual sensitivities over the egonet node set $\mathcal{V}_{\mathcal{E},i}$ into vectors, setting values to zero for elements disconnected due to a contingency:
\begin{equation}
    \nu_{P,i} = \left[ \frac{\partial |V_i|}{\partial P_k} \right]_{k \in \mathcal{V}_{\mathcal{E},i}}, \quad \nu_{Q,i} = \left[ \frac{\partial |V_i|}{\partial Q_k} \right]_{k \in \mathcal{V}_{\mathcal{E},i}}
\end{equation}
\begin{equation}
    \tau_{P,i} = \left[ \frac{\partial \theta_i}{\partial P_k} \right]_{k \in \mathcal{V}_{\mathcal{E},i}}, \quad \tau_{Q,i} = \left[ \frac{\partial \theta_i}{\partial Q_k} \right]_{k \in \mathcal{V}_{\mathcal{E},i}}
\end{equation}

The complete localized feature vector $X_i$ for bus $i$ is formulated by concatenating the local demand features, the egonet topological degree, the raw branch admittances, and the localized physical sensitivity vectors. This vector serves as the input to the predictive models is given as:
\begin{equation}
\label{eq:inputs}
\begin{split}
    X_i = \bigg[ & P_i, Q_i, \sum_{j \in \mathcal{N}_i} P_j, \sum_{j \in \mathcal{N}_i} Q_j, |\mathcal{N}_i|, \\
    & G_i, B_i, \nu_{P,i}, \nu_{Q,i}, \nu_{g,i}, \nu_{b,i}, \tau_{P,i}, \tau_{Q,i}, \tau_{g,i}, \tau_{b,i} \bigg]^\top
\end{split}
\end{equation}
It can be noted that the defined features can be easily updated with a change in the topology, for example, if there is a line contingency or network reconfiguration, the feature for the connected bus-level GPs needs to be updated and re-trained instead of re-training for the whole network. This is a major advantage of the proposed scheme, making it scalable and topology-adaptive.

\subsubsection{5-bus example network}
\label{sec:5bus_example}
In order to illustrate the modular bus-level GP structure and associated modified egonet definition for input feature extraction, we show an example for a small 5-bus network.
Figure~\ref{fig:illustrative_example_1} shows a 5-bus network, while Figure \ref{fig:illustrative_example_2} illustrates the bus-level egonet definition for target bus $V_2$ under normal operation (left) and under N$-1$ contingency (right). Under normal operation, $V_2$ connects to neighbors $V_1$ and $V_3$. The egonet node set is $\mathcal{V}_{\mathcal{E},2} = \{1, 2, 3\}$, resulting in a node degree of $|\mathcal{N}_2| = 2$. The expanded egonet branch set incorporates the direct connections and the adjacent lines routed through the neighbors, yielding $\mathcal{L}_{\mathcal{E},2} = \{(1,2), (2,3), (3,4)\}$. The aggregated power injections for the neighborhood are $P_1 + P_3$ and $Q_1 + Q_3$. The raw branch admittances and the localized voltage magnitude\footnote{The features of voltage angle sensitivites are not shown for brevity.} sensitivity vectors for active power, reactive power, conductance, and susceptance are computed over these sets:
\begin{figure}[t]
\vspace{-1em}
    \centering
    \begin{subfigure}[b]{\columnwidth}
        \centering
        \includegraphics[width=0.57\columnwidth]{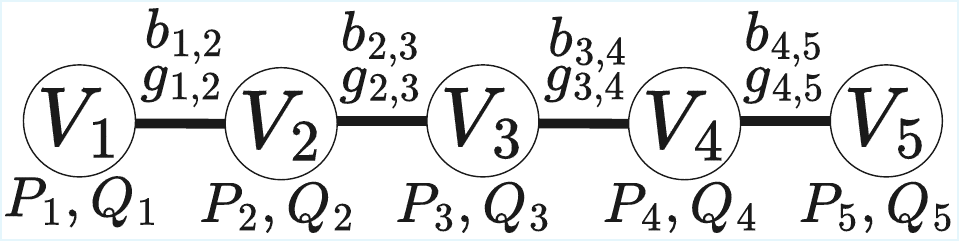}
        \caption{Original 5-bus test case.}
        \label{fig:illustrative_example_1}
    \end{subfigure}
    \hfill
    \begin{subfigure}[b]{\columnwidth}
        \centering
        \includegraphics[width=0.65\columnwidth]{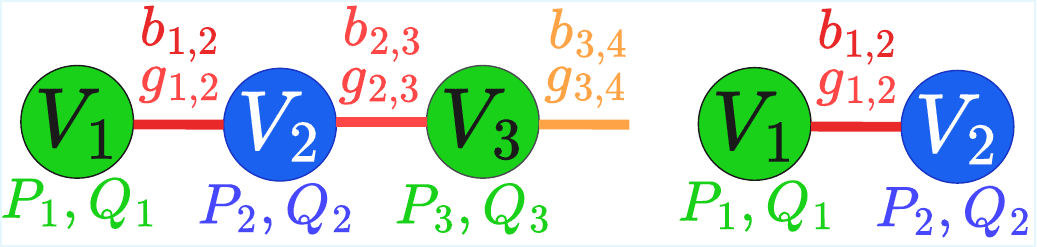}
        \caption{$V_2$ egonet under normal (left) and a contingency on line 2-3 (right).}
        \label{fig:illustrative_example_2}
    \end{subfigure}
    \hfill
    \begin{subfigure}[b]{\columnwidth}
        \centering
        \includegraphics[width=0.8\columnwidth]{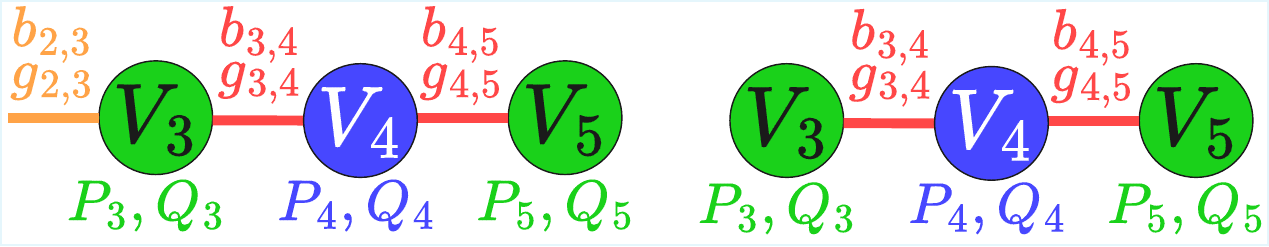}
        \caption{$V_4$ egonet under normal (left) and a contingency on line 2-3 (right).}
        \label{fig:illustrative_example_3}
    \end{subfigure}
    \caption{Illustrative example of the modified egonet definition over a 5-bus system. The subfigures display (a) the original grid, (b) the egonet for target node $V_2$, and (c) the egonet for target node $V_4$. The target node is blue, one-hop neighbors are green, direct connecting lines are red, and lines connecting neighbors to external nodes are orange. The extracted features adapt to topological changes, as demonstrated by the transition from normal operation to an N$-1$ contingency on line 2-3.}
    \label{fig:illustrative_example_complete}
\end{figure}



\begin{subequations}
\begin{align}
    G_2 = \left[ g_{1,2}, g_{2,3}, g_{3,4} \right]^\top, \quad B_2 = \left[ b_{1,2}, b_{2,3}, b_{3,4} \right]^\top
\end{align}
\begin{align}
    \nu_{P,2} = \left[ \frac{\partial |V_2|}{\partial P_1}, \frac{\partial |V_2|}{\partial P_2}, \frac{\partial |V_2|}{\partial P_3} \right]^\top ~
    \nu_{Q,2} = \left[ \frac{\partial |V_2|}{\partial Q_1}, \frac{\partial |V_2|}{\partial Q_2}, \frac{\partial |V_2|}{\partial Q_3} \right]^\top
\end{align}
\begin{align}
    \nu_{g,2} = \left[ \frac{\partial |V_2|}{\partial g_{1,2}}, \frac{\partial |V_2|}{\partial g_{2,3}}, \frac{\partial |V_2|}{\partial g_{3,4}} \right]^\top~
    \nu_{b,2} = \left[ \frac{\partial |V_2|}{\partial b_{1,2}}, \frac{\partial |V_2|}{\partial b_{2,3}}, \frac{\partial |V_2|}{\partial b_{3,4}} \right]^\top
\end{align}
\end{subequations}
The resulting baseline input feature vector for the Gaussian Process is defined as $X_2 = [P_2, Q_2, P_1 + P_3, Q_1 + Q_3, 2, G_2, B_2, \nu_{P,2}, \nu_{Q,2}, \nu_{g,2}, \nu_{b,2}, \tau_{P,2}, \tau_{Q,2}, \tau_{g,2}, \tau_{b,2}]^\top$.

During an N$-1$ contingency on line $(2,3)$, the grid topology changes. Line $(2,3)$ disconnects, and the neighborhood of $V_2$ updates to $\mathcal{N}_2' = \{1\}$, modifying the node set to $\mathcal{V}_{\mathcal{E},2}' = \{1, 2\}$ and the node degree to $|\mathcal{N}_2'| = 1$. The updated branch set reduces to $\mathcal{L}_{\mathcal{E},2}' = \{(1,2)\}$. The operational features recalculate over these restricted sets, and to maintain constant input dimensions for the GP, the raw admittances and physical sensitivities for disconnected elements are set to zero:
\begin{subequations}
\begin{equation}
    G_2' = \left[ g_{1,2}, 0, 0 \right]^\top, \quad B_2' = \left[ b_{1,2}, 0, 0 \right]^\top
\end{equation}
\begin{align}
    \nu_{P,2}' = \left[ \frac{\partial |V_2|}{\partial P_1}, \frac{\partial |V_2|}{\partial P_2}, 0 \right]^\top~
    \nu_{Q,2}' = \left[ \frac{\partial |V_2|}{\partial Q_1}, \frac{\partial |V_2|}{\partial Q_2}, 0 \right]^\top
\end{align}
\begin{equation}
    \nu_{g,2}' = \left[ \frac{\partial |V_2|}{\partial g_{1,2}}, 0, 0 \right]^\top \quad
    \nu_{b,2}' = \left[ \frac{\partial |V_2|}{\partial b_{1,2}}, 0, 0 \right]^\top
\end{equation}
\end{subequations}
The post-contingency feature vector updates to $X_2' = [P_2, Q_2, P_1, Q_1, 1, G_2', B_2', \nu_{P,2}', \nu_{Q,2}', \nu_{g,2}', \nu_{b,2}', \tau_{P,2}', \tau_{Q,2}',$ $ \tau_{g,2}', \tau_{b,2}']^\top$. 
Figure \ref{fig:illustrative_example_3} illustrates the egonet for target bus $V_4$. Under normal operation, the node set is $\mathcal{V}_{\mathcal{E},4} = \{3, 4, 5\}$ and the branch set is $\mathcal{L}_{\mathcal{E},4} = \{(2,3), (3,4), (4,5)\}$. 
Similar to bus 2, the baseline feature vector is defined as $X_4 = [P_4, Q_4, P_3 + P_5, Q_3 + Q_5, 2, G_4, B_4, \nu_{P,4}, \nu_{Q,4}, \nu_{g,4}, \nu_{b,4}, \tau_{P,4}, \tau_{Q,4}, \tau_{g,4}, \tau_{b,4}]^\top$.

Following the disconnection of line $(2,3)$, the structural neighborhood of $V_4$ remains physically unchanged ($V_3$ and $V_5$), maintaining the node set $\mathcal{V}_{\mathcal{E},4}' = \{3, 4, 5\}$ and consequently the power sensitivities ($\nu_{P,4}'=\nu_{P,4}$; $\nu_{Q,4}'=\nu_{Q,4}$, $\tau_{P,4}'=\tau_{P,4}$, $\tau_{Q,4}'=\tau_{Q,4}$). The branch set updates to $\mathcal{L}_{\mathcal{E},4}' = \{(3,4), (4,5)\}$ due to the isolation of $V_3$ from $V_2$, and the admittance sensitivities ($\nu_{b,4}'$, $\nu_{g,4}'$, $\tau_{b,4}$, $\tau_{g,4}$) are updated. 
The post-contingency feature vector is defined as $X_4' = [P_4, Q_4, P_3 + P_5, Q_3 + Q_5, 2, G_4', B_4', \nu_{P,4}', \nu_{Q,4}', \nu_{g,4}', \nu_{b,4}',\tau_{P,4}', \tau_{Q,4}', \tau_{g,4}', \tau_{b,4}']^\top$.
\subsection{Random Fourier Features (RFF) Approximation}
\label{met:rff}
The application of GP regression to a large-scale power system presents two computational challenges: (i) scalability of offline training and (ii) online prediction speed. In the \textit{training phase}, the GP models are usually trained across a vast and diverse set of network configurations and load profiles \cite{yang2024topology, liu2022topology} to achieve true zero-shot topology as well as load adaptiveness. For large-scale grids, this requirement translates into massive training datasets containing thousands of samples, and standard GP training becomes computationally intractable in this regime due to the $\mathcal{O}(N^3)$ complexity required to invert the covariance matrix \cite{tan2026gaussian}. During the \textit{prediction phase}, real-time power system operations demand state predictions within milliseconds, and standard GP inference scales poorly with the number of training samples, restricting its real-time applicability. 

Although the proposed modular bus-level GP architecture in this work is highly scalable compared to existing non-modular GP learning models, we propose using Random Fourier Features (RFFs)-based GP Kernels approximation to further enhance the computational performance, which maps the training data into a lower-dimensional space.

For the input feature vector $X_i$ extracted for bus $i$ (as defined in \eqref{eq:inputs}) and a characteristic length scale parameter $l$, the RFF mapping projects the data into a lower-dimensional space $\mathbb{R}^{2D}$. A weight matrix $W_l \in \mathbb{R}^{13 \times D}$ is sampled from a normal distribution $\mathcal{N}(0, 1/l^2)$. The feature map $\phi_l(X_i)$, parameterized by the length scale $l$, is defined as:
\begin{equation}
\label{eq:rff}
\phi_l(X_i) = \frac{1}{\sqrt{D}} \left[ \cos(W_l^\top X_i)^\top, \sin(W_l^\top X_i)^\top \right]^\top
\end{equation}

Instead of utilizing features $X_i$ directly, the predictive algorithms operate on this transformed space $\phi_l(X_i)$. Through this projection, the original $\mathcal{O}(N^3)$ training complexity is reduced to $\mathcal{O}(ND^2)$ \cite{polyzos2021ensemble}. The dimension $D$ is a hyperparameter; typically, a configuration where $D \ll N$ provides sufficient approximation accuracy across the majority of tasks \cite{polyzos2021online}. This complexity reduction allows the framework to achieve faster computational time on both training and prediction phases.

\subsection{Ensemble Gaussian Process Model}
\label{met:egp}
The data-driven modeling of AC power flow is challenging due to large simultaneous variations 
in active and reactive power demands, as well as topological changes.
Consequently, a single GP model may lack the capacity to generalize across this combined operational and structural complexity, often failing to achieve true topology adaptiveness \cite{tan2026gaussian,pareek2025data}.

To overcome this limitation and to improve the model's predictive capacity, this framework replaces the single-GP approach with an ensemble of $M$ independent GPs \cite{polyzos2021ensemble,polyzos2021online}, 
called experts. Each expert $m \in \{1, \dots, M\}$ is parameterized by a unique characteristic length scale $l_m$, thus having its own specific RFF projection space $\phi_{l_m}(X_i)$ for the input features.

Such a scheme has the advantage of aggregating the predictions of multiple GPs, where the ensemble can adapt its behavior based on the specific combination of loading conditions and topological features presented in the input vector. This enhances the generalizability of the model, allowing it to achieve robust zero-shot adaptiveness across unobserved post-contingency configurations.

During the training phase, learning is executed independently for each expert. The fitness of expert $m$ to the training data is evaluated using the exact log marginal likelihood, $\mathcal{L}_m$. This metric quantifies the probability of the observed data given the hyperparameters. The contribution of each expert to the final ensemble is determined by a probability weight $w_m$, computed via a softmax function applied to the log marginal likelihoods:
\begin{equation}
\label{eq:weight_gp}
w_m = \frac{\exp(\mathcal{L}_m - \max_{k} \mathcal{L}_k)}{\sum_{j=1}^M \exp(\mathcal{L}_j - \max_{k} \mathcal{L}_k)} 
\end{equation}

During the prediction phase, the individual predictions generated by all $M$ experts are aggregated using these assigned probability weights. This weighted aggregation ensures that the framework dynamically shifts its reliance toward the specific GP expert whose length scale is best suited for the topological state currently being evaluated, yielding a highly accurate, unified power flow prediction.

\subsection{Training and Testing and Prediction Process}
\label{met:power_flow_formulation}

The modeling framework is divided into a training phase and a testing phase. During the training phase, the standardized dataset comprising $N$ observations is denoted as $\bar{X}_{\text{train},i}$. In standard GP regression, the model learns by constructing an exact $N \times N$ covariance matrix $K$. Each element of this matrix is computed by evaluating the SE kernel function over all possible pairs of training inputs, such that $K_{p,q} = \kappa_m(X_{p}, X_{q})$ for any $X_{p}, X_{q} \in \bar{X}_{\text{train},i}$. 

By utilizing the RFF projection detailed in Section \ref{met:rff}, the SE kernel between two inputs is approximated by the inner product of their mapped features, such that $\kappa_m(X_i, X'_i) \approx \phi_{l_m}(X_i)^\top \phi_{l_m}(X'_i)$. Consequently, projecting the entire training dataset yields a transformed feature matrix $\Phi_{l_m} \in \mathbb{R}^{N \times 2D}$, where each row corresponds to a mapped input $\phi_{l_m}(X_i)^\top$. The covariance matrix is then approximated as $K \approx \Phi_{l_m} \Phi_{l_m}^\top$ and learned during training.

With the target training matrix denoted as $Y_{\text{train},i} = [\left| V_{\text{train},i} \right|, \theta_{\text{train},i}] \in \mathbb{R}^{N \times 2}$, the GP optimizes its hyperparameters through Maximum Likelihood Estimation (MLE) on the marginal likelihood of the training data. Simultaneously, the log marginal likelihood ($\mathcal{L}_m$) for each expert's fit is used during the training phase to compute a unified probability weight $w_{i,m}$ for each bus $i$ and expert $m$ in the ensemble, as defined in \eqref{eq:weight_gp}.

During the testing phase, the model assigns a predictive mean to a new, unobserved input state $\bar{x}_{\text{test},i}$. In standard GP regression, this expected value is calculated as $\mu(x_*) = k_*^\top (K + \sigma_\epsilon^2 I)^{-1} Y$, where $k_*$ is the covariance vector between the test input and the training set. By substituting the covariance components with their respective RFF approximations ($K \approx \Phi_{l_m} \Phi_{l_m}^\top$ and $k_* \approx \Phi_{l_m} \phi_{l_m}(\bar{x}_{\text{test},i})$), the two-dimensional predictive expected value $[|\hat{V}_{m,i}|, \hat{\theta}_{m,i}]$ for the $m$-th GP expert is computed over the projected test input $\phi_{l_m}(\bar{x}_{\text{test},i})$ as:
\begin{equation}
    [|\hat{V}_{m,i}|, \hat{\theta}_{m,i}] = \phi_{l_m}(\bar{x}_{\text{test},i})^\top \left( \Phi_{l_m}^\top \Phi_{l_m} + \sigma_\epsilon^2 I \right)^{-1} \Phi_{l_m}^\top Y_{\text{train},i}
\end{equation}

Finally, applying the ensemble aggregation strategy defined in Section \ref{met:egp}, the overall topology-adaptive predictions for bus $i$ are calculated as the weighted sum of the individual predictive means from all $M$ experts:
\begin{equation}
    |\hat{V}_{i}| = \sum_{m=1}^M w_{i,m} |\hat{V}_{m,i}|  \quad , \quad   \hat{\theta}_{i} = \sum_{m=1}^M w_{i,m} \hat{\theta}_{m,i}
\end{equation}

\section{Main results}
\label{sec:main_results}

\subsection{Simulation Setup}
\label{subsec:simulation_setup}

To simulate varying operating conditions, the active and reactive power loads at each bus were uniformly perturbed by $\pm 50\%$ from their nominal base values. The dataset was partitioned to evaluate model performance on both seen and unseen network topologies. The sampling and splitting criteria are defined as follows:
\begin{itemize}
    \item \textbf{N-0 and N-1 topologies:} 100 operating points were generated for the base topology and each individual single-line outage. This subset was split into $60\%$ for training and $40\%$ for testing.
    \item \textbf{N-2 topologies:} 500 operating points were generated by randomly sampling simultaneous two-line outages. These samples were allocated exclusively to the \textit{testing} set.
    \item \textbf{N-3 topologies:} 1000 operating points were generated by randomly sampling simultaneous three-line outages, also allocated exclusively to the \textit{testing} set.
    \item \textbf{Network Reconfiguration:} To evaluate topology adaptiveness beyond line outages, $20\%$ of the transmission lines were considered switchable. For each operating point, two random lines are turned off but other two lines gets switched on, and active and reactive power loads were randomly perturbed. For this, 500 operating points were generated and allocated to the \textit{testing} set.
\end{itemize}

Any sampled topology that resulted in disconnected network components or non-converging AC power flow solutions during the training phase was discarded. The proposed methodology performance is evaluated on standard IEEE test cases (14-bus, 30-bus, 57-bus, 118-bus and 300-bus) and the PEGASE 1354  test case, which contains 1,354 buses, 260 generators, and 1,991 branches, representing a subset of the European high-voltage transmission network.

We generated the AC power flow dataset using the \texttt{pandapower} Python package. To implement the ensemble Gaussian Process methodology, we utilized the \texttt{scikit-learn} Python library. Specifically, we performed the regression tasks via the BayesianRidge and MultiOutputRegressor modules, integrating them with custom projections for RFF approximations. For all simulations, we extract $D=50$ components for RFF reduction and deploy an ensemble of 8 GPs with length scales $[0.2, 0.5, 0.8, 1.0, 1.5, 2.0, 3.0, 5.0]$.

All simulations, feature extraction routines, and model training were executed sequentially on a CPU environment using Python on a Linux 6.1 operating system, utilizing a dedicated machine equipped with 16 GB of RAM (2200 MHz) and a 12-core Intel i7-1260P processor.

To evaluate the performance of the proposed framework, its predictive accuracy is benchmarked against existing methods in the literature: PINN EVGNN \cite{nakiganda2023graph}, DeepOPF-FT\footnote{Code available at \url{https://github.com/Mzhou-cityu/DeepOPF-FT}} \cite{zhou2022deepopf}, MT-VDK \cite{pareek2025data}, TTF Ensemble \cite{jia2024two} and PG-GCNN \cite{gao2023physics}. Some of these baseline models were originally formulated for Optimal Power Flow (OPF). To ensure a fair and consistent comparison with the proposed Power Flow (PF) framework, the output layers and loss functions of the OPF-based baselines were modified to isolate and evaluate exclusively the voltage magnitude and angle predictions. All models were evaluated using the identical training and testing data split detailed in Section \ref{subsec:simulation_setup}. Furthermore, for benchmark methodologies designed to achieve topology adaptiveness through the incorporation of additional post-contingency training samples, the training procedures were restricted to enforce true zero-shot generalizability. No additional samples from unseen topologies were provided to the models during training, ensuring a fair comparison aligned with the topology-adaptiveness of the proposed work.

\subsection{Voltage Predictions under $N-k$ Contingencies}
\label{subsec:voltage_predictions}

Table~\ref{tab:main_results_rmse} presents the Root Mean Square Error (RMSE) for the voltage magnitude ($|V|$) and angle ($\theta$) predictions over normal operation (N-0) as well as N-1, N-2 and N-3 contingencies. The voltage magnitude RMSE remains below 0.005 p.u. across all evaluated test systems and contingency levels. Voltage angle errors exhibit a similar scaling trend with both contingency severity and system size, reaching a maximum RMSE of 4.87 degrees for the 300-bus network under N$-3$ conditions.

\begin{figure}[!htbp]
    \centering
    
    \begin{subfigure}[b]{\columnwidth}
        \centering
        \includegraphics[width=\columnwidth]{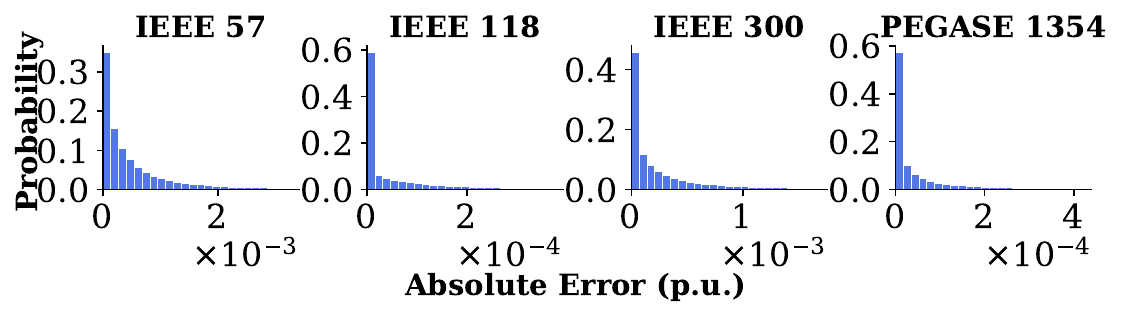}
        \caption{Voltage magnitude RMSE distribution}
        \label{fig:rmse_distribution_vm}
    \end{subfigure}

    \begin{subfigure}[b]{\columnwidth}
        \centering
        \includegraphics[width=\columnwidth]{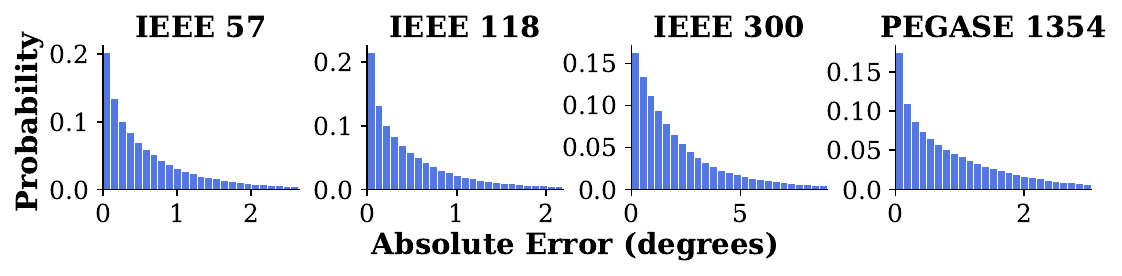}
        \caption{Voltage angle RMSE distribution}
        \label{fig:rmse_distribution_v1}
    \end{subfigure}

    \caption{Voltage magnitude (a) and angle (b) prediction errors distribution for the proposed framework over multiple test cases. Results are evaluated under unobserved N$-1$, N$-2$, and N$-3$ contingencies.}
    \label{fig:error_distribution_all}
\end{figure}

\begin{table}[!htbp]
\vspace{0.5em}
\centering
\caption{Proposed framework RMSE for Voltage Magnitude and Angle Predictions over N-0, N-1, N-2 and N-3 contingencies across different Test Systems.}
\label{tab:main_results_rmse}
\resizebox{\columnwidth}{!}{
\begin{tabular}{l cccc cccc}
\hline \hline
& \multicolumn{4}{c}{\textbf{$|V|$ RMSE (p.u.)}} & \multicolumn{4}{c}{\textbf{$\theta$ RMSE (deg.)}} \\ 
\cmidrule(lr){2-5} \cmidrule(lr){6-9}
\textbf{System} & \textbf{N-0} & \textbf{N-1} & \textbf{N-2} & \textbf{N-3} & \textbf{N-0} & \textbf{N-1} & \textbf{N-2} & \textbf{N-3} \\
\hline
IEEE 57     & 0.0005 & 0.0010 & 0.0024 & 0.0050 & 0.5493 & 0.7758 & 1.3091 & 1.7417 \\
IEEE 118    & 0.0001 & 0.0002 & 0.0003 & 0.0005 & 0.5718 & 0.7448 & 1.0228 & 1.3794 \\
IEEE 300    & 0.0007 & 0.0009 & 0.0012 & 0.0015 & 3.3901 & 3.7682 & 4.5604 & 4.8753 \\
PEGASE 1354 & 0.0002 & 0.0002 & 0.0003 & 0.0003 & 0.7881 & 1.1158 & 1.2182 & 1.2801 \\
\hline \hline
\end{tabular}
}
\end{table}
Figure~\ref{fig:error_distribution_all} details the probability distribution of the absolute prediction errors for voltage magnitudes and angles. The distributions are right-skewed and heavily concentrated near zero across the four test systems. The nodal voltage predictions don't deviate from the true AC power flow solutions, with errors of less than $2 \times 10^{-3}$ for magnitude and $5$ degrees for voltage angle.

\begin{figure}[!htbp]
    \centering
    \begin{subfigure}[b]{\columnwidth}
        \centering
        \includegraphics[width=\columnwidth]{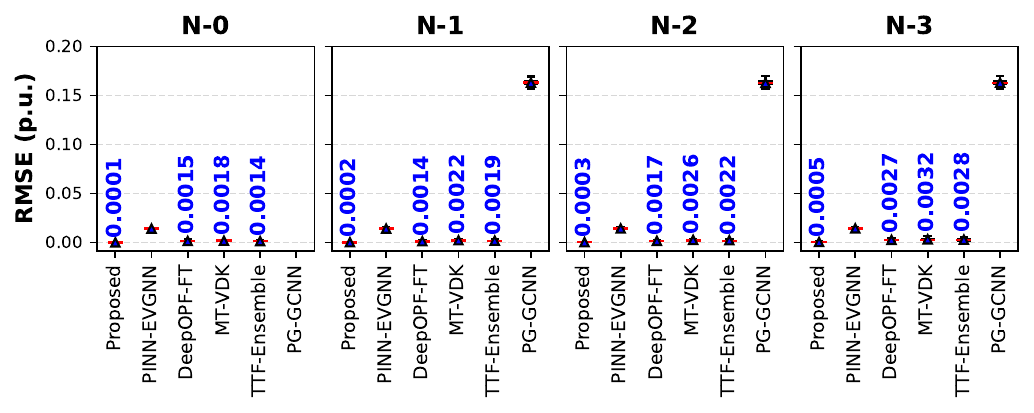}
        \caption{Voltage magnitude ($|V|$)}
        \label{fig:rmse_case14}
    \end{subfigure}
    \begin{subfigure}[b]{\columnwidth}
        \centering
        \includegraphics[width=\columnwidth]{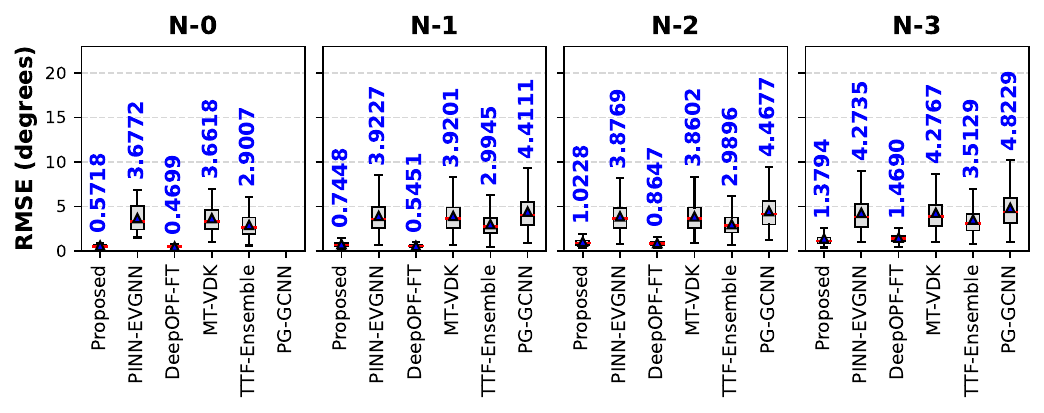}
        \caption{Voltage angle ($\theta$)}
        \label{fig:rmse_case118}
    \end{subfigure}
    \caption{Voltage magnitude (a) and angle (b) prediction RMSE across N-1, N-2, and N-3 contingencies for IEEE 118 test case.}
    \label{fig:rmse_comparison_all_cases}
\end{figure}

Figure \ref{fig:rmse_comparison_all_cases} compares the proposed work against the state-of-the-art in terms of voltage magnitude and angle RMSE across N-0 (normal operation) as well as N-1, N-2, and N-3 contingencies for the IEEE 118 test case. The proposed work achieves the lowest RMSE for voltage magnitude, with the difference increasing with the contingency, achieving 0.0005 compared to 0.0027 for the top 2 over $N-3$ contingencies. Additionally, it ranks in the top 2 for voltage angle up to $N-2$ contingencies, but goes up to top 1 for $N-3$ ($1.3794^\circ$ against $1.4690^\circ$ from the top 2).

To assess the methodologies on a larger scale, the IEEE 300-bus test case was also investigated. Only the state-of-the-art approaches capable of scaling (DeepOPF-FT and TTF-Ensemble) were evaluated alongside the proposed framework. The results are shown in Table \ref{tab:rmse_case300}. The proposed work achieves the best predictive performance across all normal and contingency scenarios. The performance gap becomes particularly pronounced during N-3 contingencies, where the proposed framework reduces the prediction error by $90.06\%$ (from 0.0161 p.u. to 0.0016 p.u) for voltage magnitude and $72.33\%$ (from $17.60^\circ$ to $4.87^\circ$) for voltage angle when compared to the second-best approach.

\begin{table}[!htbp]
    \centering
    \vspace{0.5em}
    \caption{Prediction RMSE for Voltage Magnitude and Angle on the IEEE 300-Bus Test Case}
    \label{tab:rmse_case300}
    \begin{tabular}{@{}llcccc@{}}
        \toprule
        \multirow{2}{*}{\textbf{Metric}} & \multirow{2}{*}{\textbf{Approach}} & \multicolumn{4}{c}{\textbf{Topologies}} \\ \cmidrule(l){3-6} 
        & & \textbf{N-0} & \textbf{N-1} & \textbf{N-2} & \textbf{N-3} \\ 
        \midrule
        \multirow{3}{*}{\shortstack{$|V|$ RMSE \\ (p.u.)}} 
        & Proposed & \textbf{0.0005} & \textbf{0.0008} & \textbf{0.0011} & \textbf{0.0016} \\
        & DeepOPF-FT & 0.0101 & 0.0136 & 0.0158 & 0.0161 \\
        & TTF-Ensemble & 0.0150 & 0.0173 & 0.0187 & 0.0178 \\ 
        \midrule
        \multirow{3}{*}{\shortstack{$\theta$ RMSE \\ (deg.)}} 
        & Proposed & \textbf{3.3901} & \textbf{3.7681} & \textbf{4.5603} & \textbf{4.8753} \\
        & DeepOPF-FT & 8.9782 & 13.5885 & 15.1938 & 17.6078 \\
        & TTF-Ensemble & 22.797 & 29.7926 & 30.6417 & 29.6091 \\ 
        \bottomrule
    \end{tabular}
\end{table}

\begin{figure*}[!htbp]
    \centering
    \begin{subfigure}[b]{0.24\textwidth}
        \centering
        \includegraphics[width=\textwidth]{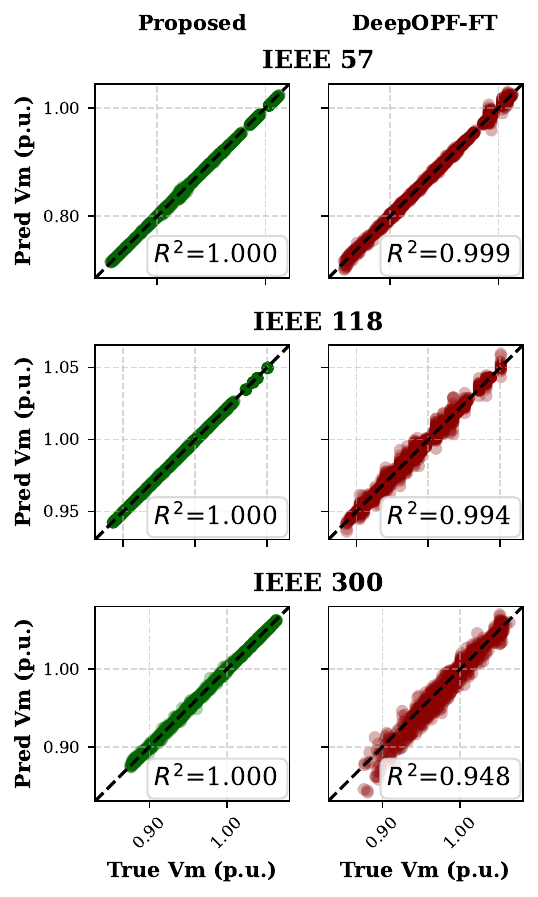}
        \caption{N-0 (normal operation)}
        \label{fig:scatter_n0}
    \end{subfigure}
    \hfill
    \begin{subfigure}[b]{0.24\textwidth}
        \centering
        \includegraphics[width=\textwidth]{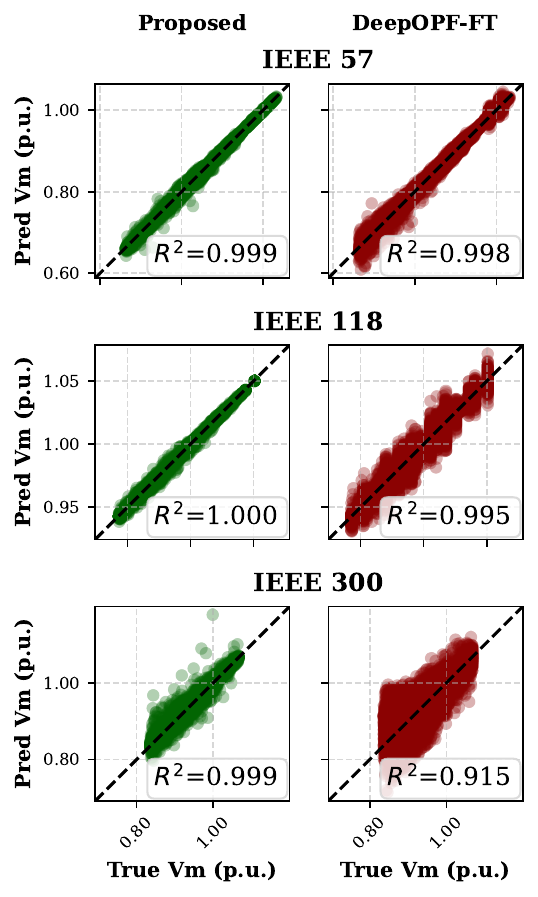}
        \caption{N-1 contingencies}
        \label{fig:scatter_n1}
    \end{subfigure}
    \hfill
    \begin{subfigure}[b]{0.24\textwidth}
        \centering
        \includegraphics[width=\textwidth]{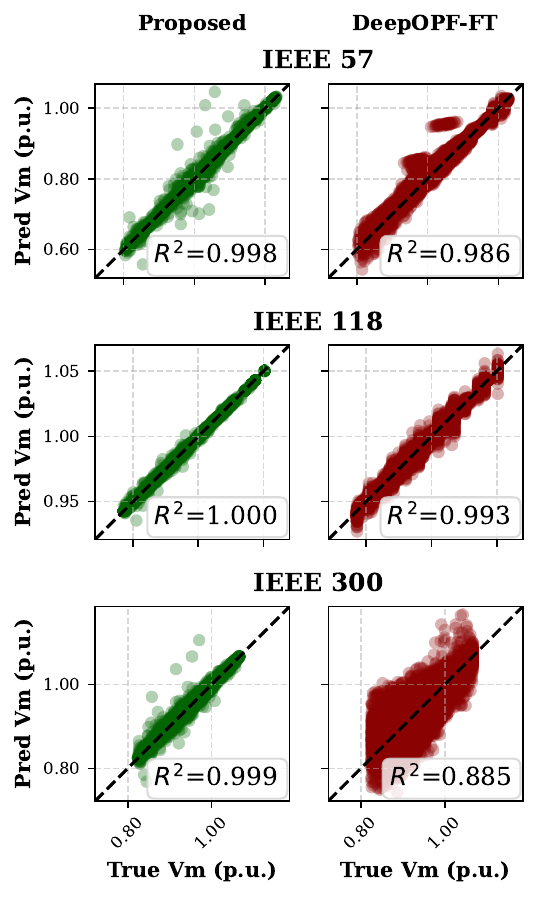}
        \caption{N-2 contingencies}
        \label{fig:scatter_n2}
    \end{subfigure}
    \hfill
    \begin{subfigure}[b]{0.24\textwidth}
        \centering
        \includegraphics[width=\textwidth]{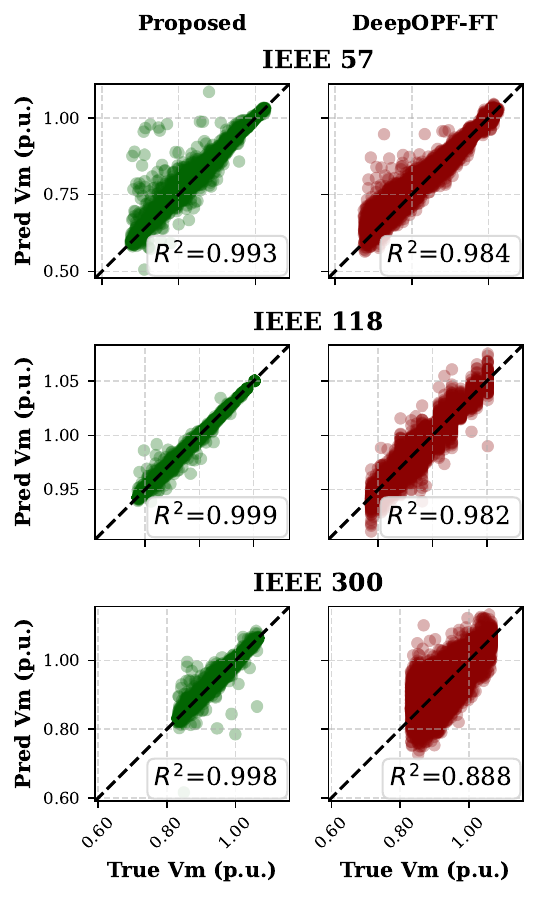}
        \caption{N-3 contingencies}
        \label{fig:scatter_n3}
    \end{subfigure}
    \caption{True vs. Predicted voltage magnitude ($|V|$, in p.u.) comparison between the Proposed and DeepOPF-FT methodologies under unseen N-0, N-1, N-2 and N-3 contingencies.}
    \label{fig:comparison_scatter_all}
    \vspace{-1em}
\end{figure*}

Prediction quality was further analyzed by comparing the proposed approach against the DeepOPF-FT model \cite{zhou2022deepopf}, which exchibited the best performance from the literature. Figure \ref{fig:comparison_scatter_all} shows the comparison for voltage magnitude ($|V|$). The DeepOPF-FT model demonstrates competitive performance on smaller systems (IEEE 57 and 118), but on the larger IEEE 300 system, performance degrades as the contingency severity increases. While its predictions are good under normal (N$-0$) conditions, the scatter plots for N$-1$, N$-2$, and especially N$-3$ contingencies show wider dispersion. Specially for the IEEE 300-bus system, where the $R^2$ score drops to 0.888 under N$-3$ events, compared with 0.998 from the proposed work.

\begin{figure*}[!htbp]
    \centering
    \begin{subfigure}[b]{0.24\textwidth}
        \centering
        \includegraphics[width=\textwidth]{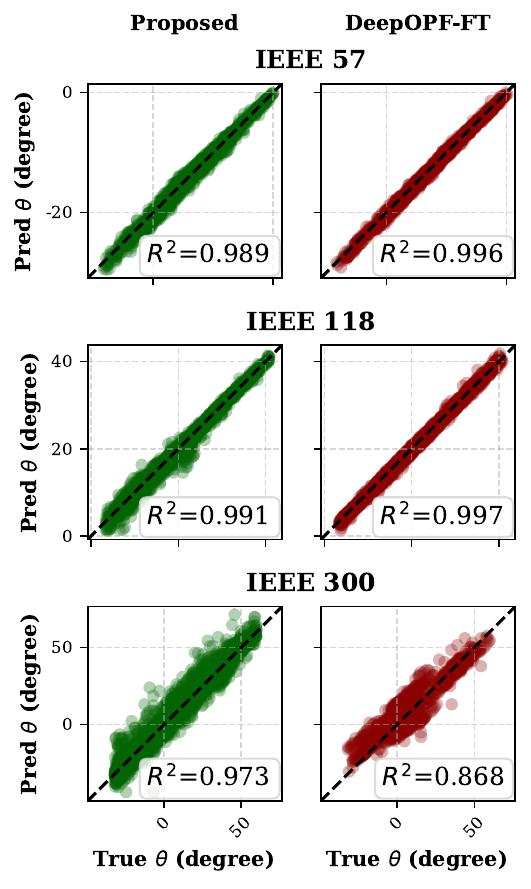}
        \caption{N-0 (normal operation)}
        \label{fig:scatter_n0}
    \end{subfigure}
    \hfill
    \begin{subfigure}[b]{0.24\textwidth}
        \centering
        \includegraphics[width=\textwidth]{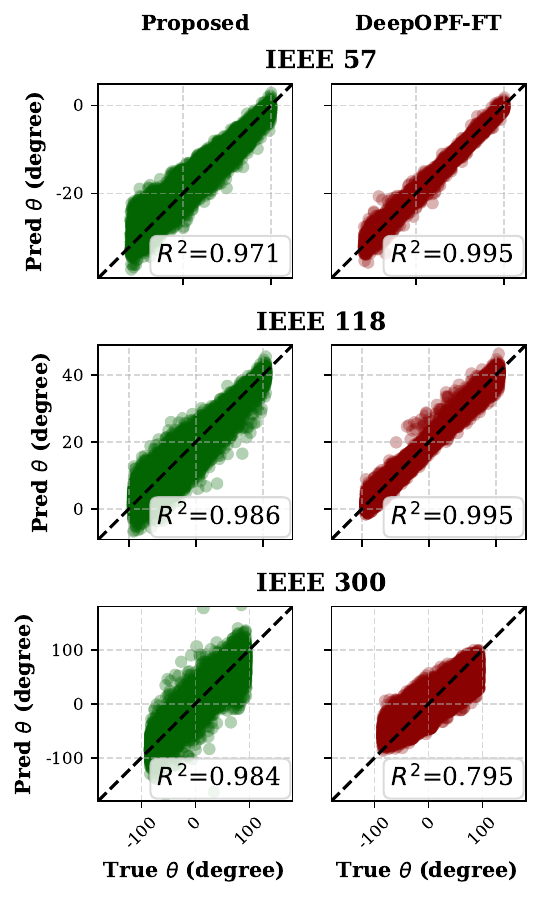}
        \caption{N-1 contingencies}
        \label{fig:scatter_n1}
    \end{subfigure}
    \hfill
    \begin{subfigure}[b]{0.24\textwidth}
        \centering
        \includegraphics[width=\textwidth]{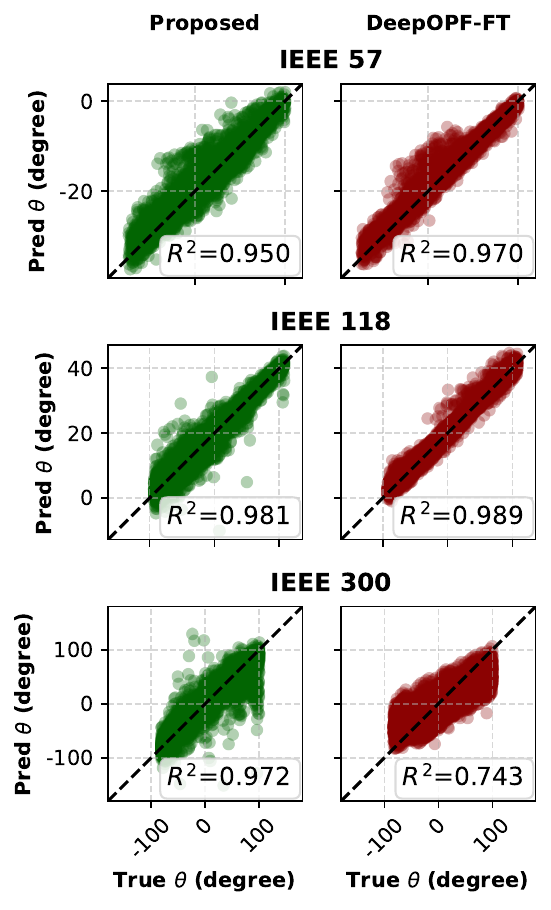}
        \caption{N-2 contingencies}
        \label{fig:scatter_n2}
    \end{subfigure}
    \hfill
    \begin{subfigure}[b]{0.24\textwidth}
        \centering
        \includegraphics[width=\textwidth]{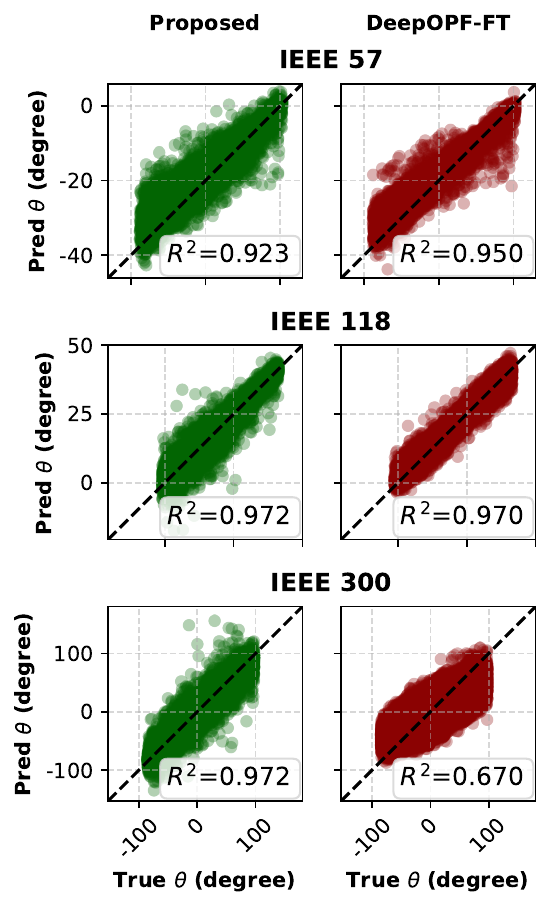}
        \caption{N-3 contingencies}
        \label{fig:scatter_n3}
    \end{subfigure}
    \caption{True vs. Predicted voltage angle ($\theta$, in degrees) comparison between the Proposed and DeepOPF-FT methodologies under unseen N-0, N-1, N-2 and N-3 contingencies.}
    \label{fig:comparison_scatter_all_va}
    \vspace{-1.5em}
\end{figure*}

Figure \ref{fig:comparison_scatter_all_va} shows the prediction quality for voltage angles ($\theta$). The baseline DeepOPF-FT model demonstrates strong predictive performance on the smaller IEEE 57- and 118-test cases. For the larger IEEE 300 system, the DeepOPF-FT decreases in performance, with $R^2$ scores dropping from 0.868 under normal operation to 0.670 under N-3 contingencies. On the other hand, the proposed work maintains good predictive performance on larger networks, outperforming DeepOPF-FT on the 300-bus network and achieving an $R^2$ of 0.972 even under N-3 contingencies, compared to the baseline's 0.670. 

The proposed model exhibits slightly lower performance, with wider dispersion, in predicting voltage angles compared to voltage magnitudes. This behavior reflects a limitation of local models versus global models, as localized learning does not fully capture system-wide angular references and patterns. Nevertheless, as shown in Table \ref{tab:rmse_case300}, the proposed framework still outperforms existing methods by achieving consistently lower RMSE. Furthermore, this localized approach provides a significant advantage by enabling much greater scalability and speed in large power systems, as shown later in Table~\ref{tab:comp_time_proposed}.

\subsection{Detection of Contingency Scenarios}
\label{subsec:contingency_scenarios}
The ability of the framework to identify contingent grid states is evaluated through the recall rate for voltage limit violations. Recall is calculated as the ratio of correctly predicted violations to the total number of actual violations present in the AC power flow reference:
$$\text{Recall} = \frac{TP}{TP + FN}$$
where $TP$ (True Positives) represents the number of voltage violations correctly identified by the model, and $FN$ (False Negatives) represents the number of actual violations that the model failed to detect.  Two distinct event categories are defined based on the nodal voltage magnitude $|V|$: Under-voltage scenarios ($|V| < 0.9$ p.u.) and Over-voltage scenarios ($|V| > 1.1$ p.u).

\begin{figure}[!htbp]
    \vspace{-1em}
    \centering
    \begin{subfigure}[b]{0.32\linewidth}
        \centering
        \includegraphics[width=\linewidth]{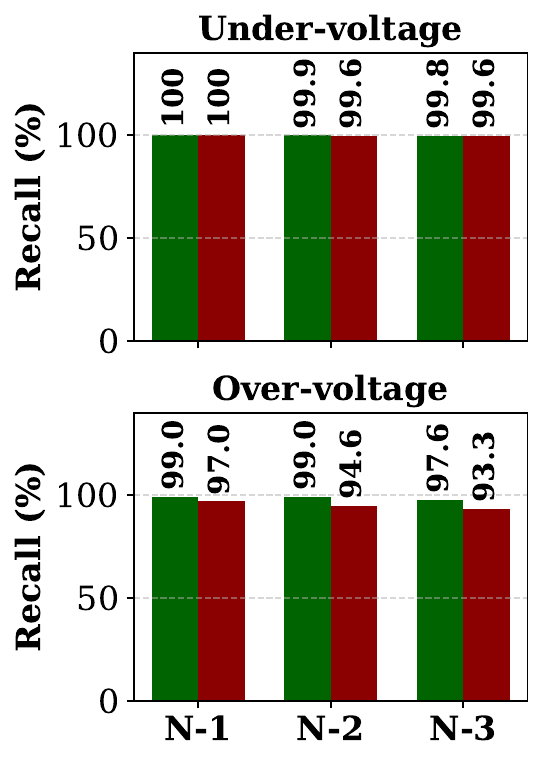}
        \caption{IEEE 57}
        \label{fig:recall_case57}
    \end{subfigure}
    \hfill
    \begin{subfigure}[b]{0.32\linewidth}
        \centering
        \includegraphics[width=\linewidth]{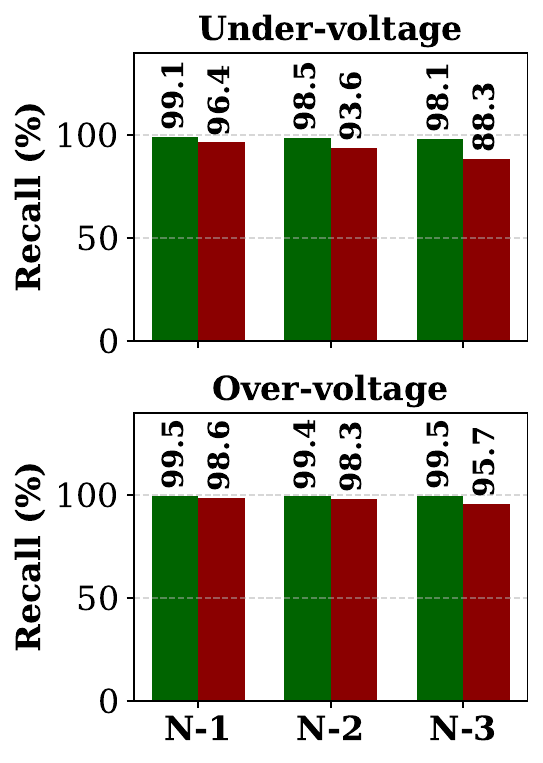}
        \caption{IEEE 118}
        \label{fig:recall_case118}
    \end{subfigure}
    \hfill
    \begin{subfigure}[b]{0.32\linewidth}
        \centering
        \includegraphics[width=\linewidth]{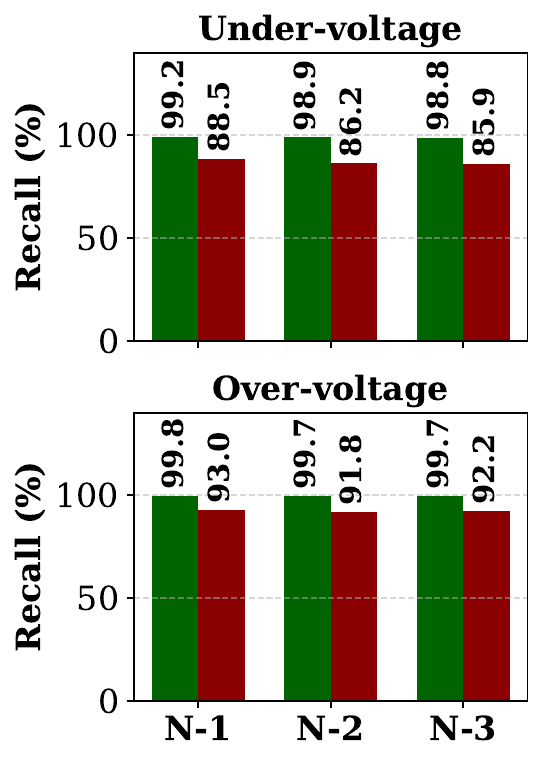}
        \caption{IEEE 300}
        \label{fig:recall_case300}
    \end{subfigure}
    \caption{Recall rates for detecting voltage violations. The proposed work (green) is compared against the DeepOPF-FT baseline (red) under unobserved N-1, N-2, and N-3 contingencies.}
    \label{fig:comparison_recall_all}
\end{figure}

Figure~\ref{fig:comparison_recall_all} shows the recall rates across the IEEE 57, 118, and 300-bus systems for N$-1$, N$-2$, and N$-3$ contingencies. We compare the proposed approach with DeepOPF-FT \cite{zhou2022deepopf}. The proposed work (green) maintains a recall rate of 98\% for both under-voltage and over-voltage detection, even on N-$3$ contingencies. In contrast, the baseline (red) recall reduces as system complexity and contingency severity increase.

\begin{figure}[!htbp]

    \centering
    \begin{subfigure}[b]{0.49\linewidth}
        \centering
        \includegraphics[width=\linewidth]{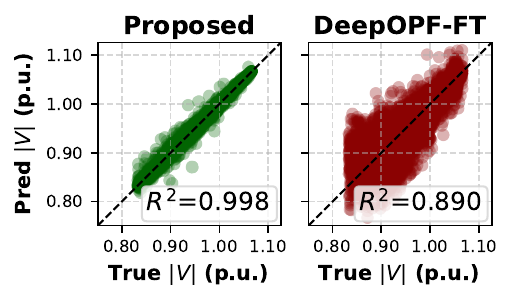}
        \caption{Voltage magnitude}
        \label{fig:reconfiguration_rmse_vm}
    \end{subfigure}
    \hfill
    \begin{subfigure}[b]{0.49\linewidth}
        \centering
        \includegraphics[width=\linewidth]{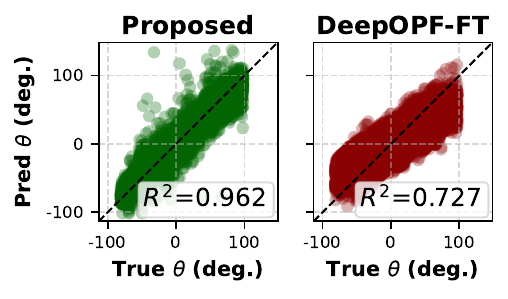}
        \caption{Voltage angle}
        \label{fig:reconfiguration_rmse_va}
    \end{subfigure}
    \caption{Voltage magnitude (a) and angle (b) prediction performance across network reconfigurations for the IEEE 300 test case.}
    \label{fig:rmse_comparison_reconfiguration}
\end{figure}
\subsection{Voltage Prediction under Network Reconfiguration}
We also assess the prediction performance under network reconfiguration. Figure \ref{fig:rmse_comparison_reconfiguration} shows the prediction performance of the proposed methodology and the DeepOPF-FT baseline under network reconfiguration scenarios for the IEEE 300-bus system. The proposed work achieves high prediction performance, outperforming the baseline in both voltage magnitude and angle predictions. Regarding RMSE evaluation, for voltage magnitude the proposed approach achieves an RMSE of 0.00185 p.u, against 0.01456 from DeepOPF-FT, a $87.29\%$ reduction. For voltage angle, the proposed model yields an RMSE of 6.63 degrees, compared to 17.70 degrees from the baseline, a $62.5\%$ reduction.

\subsection{Ablation Study}
\label{subsec:ablation_study}
An ablation study is conducted on the IEEE 300-bus system to evaluate the contributions of three components of the proposed work to prediction performance: the localized admittance sensitivities, the Random Fourier Features (RFF) approximation, and the ensemble of Gaussian Processes. 

Table \ref{tab:ablation_case57} reports performance metrics for all combinations of the three components, demonstrating that the proposed methodology with all three components improves prediction accuracy across all contingencies.

\begin{table}[!htbp]
\centering
\vspace{0.5em}
\caption{Impact of Model Components on Accuracy for the IEEE 300 System. The evaluated components include localized admittance sensitivities (Sens.), Random Fourier Features approximation (RFF), and the ensemble of Gaussian Process experts (Ens.).}
\label{tab:ablation_case57}
\resizebox{\columnwidth}{!}{
\begin{tabular}{l ccc ccc ccc}
\hline \hline
& \multicolumn{3}{c}{\textbf{Components}} & \multicolumn{3}{c}{\textbf{RMSE $|V|$ (p.u.)}} & \multicolumn{3}{c}{\textbf{RMSE $\theta$ (deg.)}} \\ 
\cmidrule(lr){2-4} \cmidrule(lr){5-7} \cmidrule(lr){8-10}
\textbf{Model} & \textbf{Sens.} & \textbf{RFF} & \textbf{Ens.} & \textbf{N-1} & \textbf{N-2} & \textbf{N-3} & \textbf{N-1} & \textbf{N-2} & \textbf{N-3} \\
\hline
$M_0$ & $\times$   & $\times$   & $\times$   & 0.0085 & 0.0122 & 0.0152 & 3.2366 & 3.7731 & 4.5430 \\
$M_1$ & $\times$   & $\times$   & \checkmark & 0.0085 & 0.0122 & 0.0152 & 3.2366 & 3.7731 & 4.5430 \\
$M_2$ & $\times$   & \checkmark & $\times$   & 0.0121 & 0.0172 & 0.0226 & 3.3085 & 3.9209 & 4.7870 \\
$M_3$ & $\times$   & \checkmark & \checkmark & 0.0074 & 0.0113 & 0.0139 & 3.0539 & 3.6406 & 4.3567 \\
$M_4$ & \checkmark & $\times$   & $\times$   & 0.0078 & 0.0112 & 0.0142 & 3.0812 & 3.6773 & 4.4933 \\
$M_5$ & \checkmark & $\times$   & \checkmark & 0.0078 & 0.0112 & 0.0142 & 3.0812 & 3.6773 & 4.4933 \\
$M_6$ & \checkmark & \checkmark & $\times$   & 0.0128 & 0.0183 & 0.0239 & 3.3100 & 4.0125 & 4.8566 \\
$M_7$ & \checkmark & \checkmark & \checkmark & \textbf{0.0009} & \textbf{0.0012} & \textbf{0.0015} & \textbf{2.3932} & \textbf{2.5339} & \textbf{2.5739} \\
\hline \hline
\end{tabular}
}
\end{table}

\subsection{Scalability under larger systems}
\label{subsec:scalability_larger_systems}

\begin{figure}[!htbp]
    \centering
    \includegraphics[width=1\linewidth]{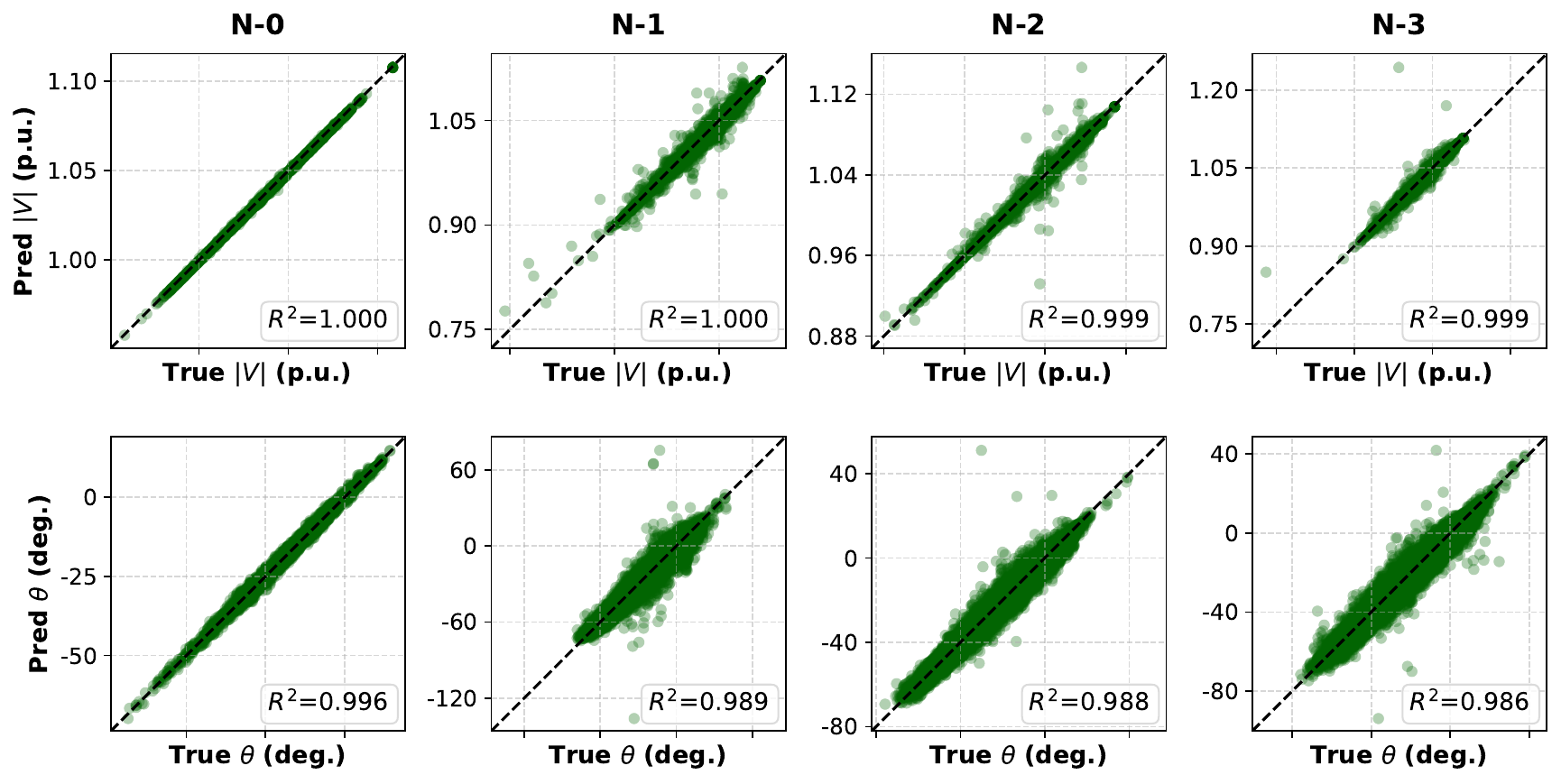}
    \caption{True versus predicted voltage magnitudes ($|V|$) and angles ($\theta$) for the PEGASE 1354 system across N-0, N-1, N-2 and N-3 contingencies.}
    \label{fig:case1354pegase}
\end{figure}

Figure \ref{fig:case1354pegase} presents the scatter plots of true versus predicted voltage magnitudes ($|V|$) and voltage angles ($\theta$) across N-0, N-1, N-2 and N-3 contingencies. The proposed approach has a $R^2$ equal to or exceeding 0.929 across all contingencies levels, in both voltage magnitude and angle. 

\subsection{Computational Performance}
\label{subsec:computational_performance}

Table \ref{tab:comp_time_proposed} details the computational times for the proposed framework\footnote{All experiments were executed using Python on a Windows operating system, utilizing a personal computer with 16 GB of RAM (2200 MHz) and a 12-core Intel i7-1260P processor.}. The application of the RFF technique reduces the \textit{training} computational complexity of the proposed framework to $\mathcal{O}(N D^2)$, where $D$ is the constant dimension of the randomized feature space and $N$ is the number of training samples. Since $D$ remains constant, the overall computational complexity of the framework reduces to $\mathcal{O}(N)$. The \textit{testing} computational complexity depends on both the total number of buses and the individual input dimension at each bus, which varies according to local branch connectivity. 


Table \ref{tab:comp_time_proposed} also compares the computational time of the proposed framework with state-of-the-art approaches, focusing on methodologies capable of evaluating the IEEE 300 test case. DeepOPF-FT \cite{zhou2022deepopf} exhibits high training times due to its global system approach. The TTF Ensemble \cite{jia2024two} demonstrates competitive training times. However, its testing time is higher because the model requires online learning with new topologies. While the existing literature has run models on systems with more than 1000 nodes, these studies typically use smaller datasets that exclude $N-k$ contingencies. Applying these existing methods to larger systems (i.e. PEGASE 1354) resulted in out-of-memory errors.

\begin{table}[!htbp]
    \centering
    \vspace{0.5em}
    \caption{Computational time of the Proposed Framework and state-of-the-art approaches across IEEE test cases. OOM denotes Out-of-Memory error.}
    \label{tab:comp_time_proposed}
    \resizebox{\columnwidth}{!}{
    \begin{tabular}{llcc}
        \toprule
        \textbf{System} & \textbf{Approach} & \textbf{Train Time (s)} & \textbf{Test Time [per sample] (ms)} \\
        \midrule
        IEEE 30 & Proposed & 4.90 & 0.300\\
        \midrule
        IEEE 57 & Proposed & 233.20 & 4.528\\
        \midrule
        \multirow{3}{*}{IEEE 118} & Proposed & 701.11 & \textbf{8.886}\\
         & DeepOPF-FT \cite{zhou2022deepopf} & 7113.81 & 110\\
         & TTF Ensemble \cite{jia2024two} & 2706.49 & 17.871\\
        \midrule
        \multirow{3}{*}{IEEE 300} & Proposed & 690.81 & \textbf{8.251}\\
         & DeepOPF-FT \cite{zhou2022deepopf} & 15717.43 & 154\\
         & TTF Ensemble \cite{jia2024two} & 5263.35 & 63.595  \\
        \midrule
        \multirow{3}{*}{PEGASE 1354} & Proposed & 5929.79 & \textbf{75.130}\\
         & DeepOPF-FT \cite{zhou2022deepopf} & {OOM} & {OOM}  \\
         & TTF Ensemble \cite{jia2024two} & {OOM} & {OOM} \\
        \bottomrule
    \end{tabular}
    }
\end{table}

\section{Conclusion}
\label{sec:conclusion}
This paper presented a scalable, fast, topology-adaptive Gaussian Process framework for learning power flow solutions under unobserved grid contingencies and topologies. We proposed a modular Gaussian Process architecture, utilizing localized first-order admittance and power injection sensitivities to embed the physical impacts of structural changes directly into the predictive models. To overcome the computational limitations of standard Gaussian Processes on large datasets, the framework employs Random Fourier Features approximations and aggregates predictions through an ensemble of experts.

Simulation results across the IEEE 30, 57, 118, 300-bus and PEGASE 1354 systems demonstrated that the proposed model predicts system states under unobserved N-1, N-2, and N-3 contingencies without requiring additional training samples. The framework achieved lower prediction errors compared to existing benchmark methodologies, exhibiting robustness under severe N-3 contingencies where baseline models experienced performance degradation. Furthermore, the model maintained high recall rates, exceeding 98\% in all scenarios, for identifying under-voltage and over-voltage violations. The ablation study confirmed that the integration of localized physical sensitivities, RFF, and the ensemble strategy balances computational efficiency with prediction accuracy. Additionally, the framework demonstrates linear computational scaling with respect to system size, having a per-scenario testing time of 73ms for the PEGASE 1354 test system.

Future research will explore the framework's extension to solve the optimal power flow (OPF) problem and optimize generator active and reactive power setpoints. Furthermore, different spatial granularities will be investigated by transitioning from bus-level models to zone-based GPs. By partitioning the network into distinct regions, inter-zone dependencies can be captured by exchanging predicted boundary states and sensitivities between adjacent models. 
\bibliographystyle{IEEEtran}
\bibliography{IEEEabrv,ref}

\end{document}